\documentclass{jors}

\definecolor{mygray}{gray}{0.6}

\usepackage{graphicx}
\usepackage[square,numbers]{natbib}
\hypersetup{
    colorlinks = true,
    citecolor = blue,
}

\begin{document}

\newcommand{\michael}[1]{\noindent{\color{green} \bf{Michael: #1}}}
\newcommand{\martin}[1]{\noindent{\color{red} \bf{Martin: #1}}}

\newcommand{\code}[1]{\texttt{#1}}

{\bf Software paper for submission to the Journal of Open Research Software} \\



\rule{\textwidth}{1pt}

\section*{(1) Overview}

\vspace{0.5cm}

\section*{Title}

Probabilistic Inference on Noisy Time Series (PINTS).

\section*{Paper Authors}

1. Clerx, Michael* ORCiD:0000-0003-4062-3061 \\
2. Robinson, Martin* ORCiD:0000-0002-1572-6782 \\
3. Lambert, Ben ORCiD:0000-0003-4274-4158 \\
4. Lei, Chon Lok ORCiD:0000-0003-0904-554X \\
5. Ghosh, Sanmitra ORCiD:0000-0002-4879-7587 \\
6. Mirams, Gary R. ORCiD:0000-0002-4569-4312 \\
7. Gavaghan, David J. ORCiD:0000-0001-8311-3200

*These authors contributed equally to this work.

\section*{Paper Author Roles and Affiliations}

1. PDRA, Department of Computer Science, University of Oxford, Wolfson Building, Parks Road, Oxford, OX1 3QD, United Kingdom. \\
2. PDRA, Department of Computer Science, University of Oxford, Wolfson Building, Parks Road, Oxford, OX1 3QD, United Kingdom. \\
3. PDRA, MRC Centre for Global Infectious Disease Analysis, School of Public Health, Imperial College London, W2 1PG, United Kingdom. \\
4. DPhil Student, Department of Computer Science, University of Oxford, Wolfson Building, Parks Road, Oxford, OX1 3QD, United Kingdom. \\
5. PDRA, Department of Computer Science, University of Oxford, Wolfson Building, Parks Road, Oxford, OX1 3QD, United Kingdom. \\
6. Research Fellow, Centre for Mathematical Medicine \& Biology, School of Mathematical Sciences, University of Nottingham, Nottingham, NG7 2RD, United Kingdom. \\
7. Professor, Department of Computer Science, University of Oxford, Wolfson Building, Parks Road, Oxford, OX1 3QD, United Kingdom.

\section*{Abstract}
Time series models are ubiquitous in science, arising in any situation where researchers seek to understand how a system's behaviour changes over time.
A key problem in time series modelling is \emph{inference}; determining properties of the underlying system based on observed time series.
For both statistical and mechanistic models, inference involves finding parameter values, or distributions of parameters values, for which model outputs are consistent with observations.
A wide variety of inference techniques are available and different approaches are suitable for different classes of problems.
This variety presents a challenge for researchers, who may not have the resources or expertise to implement and experiment with these methods.
PINTS (Probabilistic Inference on Noisy Time Series --- \url{https://github.com/pints-team/pints}) is an open-source (BSD 3-clause license) Python library that provides researchers with a broad suite of non-linear optimisation and sampling methods.
It allows users to wrap a model and data in a transparent and straightforward interface, which can then be used with custom or pre-defined error measures for optimisation, or with likelihood functions for Bayesian inference or maximum-likelihood estimation.
Derivative-free optimisation algorithms --- which work without harder-to-obtain gradient information --- are included, as well as inference algorithms such as adaptive Markov chain Monte Carlo and nested sampling which estimate distributions over parameter values.
By making these statistical techniques available in an open and easy-to-use framework, PINTS brings the power of modern statistical techniques to a wider scientific audience.

\section*{Keywords}

Time series models; non-linear optimisation; MCMC sampling; nested sampling; Bayesian inference; Python

\section*{Introduction}

Time series models are common in science, where they are used to describe the dynamics of system behaviours.
In many cases, these models are non-linear and impossible to solve analytically, so that the \emph{forward problem} (predicting the model output for a given set of parameters) is computationally hard.
For such models, there is no single method which can reliably solve the \emph{inverse problem} of estimating parameter values from a noisy time trace.
Much like there is a variety of forward models, there is a diversity of approaches for parameter inference.
Further, it is often unclear which approach to apply when, meaning that researchers are required to implement a range of methods before successfully fitting their model to data.

PINTS is a software framework that allows users to easily trial and apply different inference methods to their problem.
The inference methods supplied by PINTS fall into two broad categories: \emph{optimisers}, which attempt to find a single best parameter vector, and \emph{samplers}, which aim to estimate a probability distribution over parameter values that are compatible with observed results.
Users are expected to already have a forward model (for example, a simulation) at their disposal, which they make available to PINTS by writing a simple Python wrapper.
They then define a \code{Problem} (a forward model plus a data set), on which either an \code{ErrorMeasure} (for optimisation) or a \code{LogPDF} (for optimisation and sampling) is defined.
Currently available optimisers include CMA-ES \cite{hansen2003reducing}, XNES \cite{glasmachers2010exponential}, SNES \cite{schaul2011high}, and Particle Swarm Optimisation (PSO) \cite{kennedy2011particle}.
Sampling methods include, amongst other routines, Random Walk Metropolis Markov chain Monte Carlo (MCMC) \cite{metropolis1953equation,lambert2018Student}, adaptive covariance MCMC \cite{johnstone2016uncertainty} and Population MCMC \cite{jasra2007population}.
In addition, ellipsoidal \cite{mukherjee2006nested} and rejection nested samplers \cite{skilling2006nested} are provided.
Convenience plotting methods are provided to quickly visualise the results, as well as diagnostic tools to inspect the validity of the results.

PINTS was developed as a community effort by researchers in electrochemistry, cardiac electrophysiology, and statistics, to compare different methods for solving inverse problems in a common framework.
It features a clean and transparent object-orientated API that is designed to accommodate easily new error measures, log-likelihoods, optimisers and samplers, allowing users to utilise pre-built components as much as possible, while adding their own code for problem-specific areas.
The PINTS team aims for full test coverage, and includes unit testing and extended statistical tests to verify the correct operation of all methods.

Early research using PINTS in electrochemistry has included fitting a differential-algebraic equation (DAE) model of reduction-oxidation to voltammetry measurements of a Polyoxometalates molecule \cite{pom_paper}, and the design and application of a custom hierarchical statistical model for repeat voltammetry experiments of a Ferricynide process \cite{robinson_bond_simonov_zhang_gavaghan_2018}.

Optimisation algorithms are implemented in many different software packages, (see, for example, the Python \code{scipy.optimize} module), but are often biased towards gradient-based methods, which can perform poorly for many ordinary and partial differential equations used in time series modelling.
PINTS therefore focuses on derivative-free optimisers, although we plan to add gradient-based methods for comparison.
In contrast with more general-purpose optimisation software, PINTS contains a number of error measures specifically suited to time series models, and adds the ability to use any PINTS log-likelihood class as an error measure in order to perform maximum likelihood estimation (MLE).

Dakota \cite{adam2015dakota} is a widely regarded package for parameter fitting and uncertainty quantification and is most similar to PINTS in that it offers a generic interface to call an (assumed expensive) model, as well as a wide variety of optimisers and samplers.
In contrast to the PINTS Python API, Dakota uses either a C++ or file input/output process for communication between user models and the library, and does not provide options for specifying either the error measures or the log-likelihoods of the inverse problem.
However, Dakota has some features not yet available in PINTS, such as the option to train a surrogate model, useful for very expensive model evaluations.

Other software packages that enable parameter inference and sampling for ODE models include BioBayes \cite{vyshemirsky2008biobayes}, ABC-SysBio \cite{liepe2010abc}, SYSBIONS \cite{johnson2014sysbions} and Stan \cite{carpenter2017stan}.
These packages use either a common model description format (for example, SBML) or their own language (for example, Stan's probabilistic programming language) to specify the model, presenting additional learning hurdles for a user and often restricting the class of models which can be fit.
By contrast, PINTS aims to be as general as possible to support a wider variety of models (for example, PDEs).
PyMC3 \cite{salvatier2016probabilistic} does provide a similar generic model interface to PINTS but, as with the other packages, specialises in one sampling method, whereas PINTS aims to support a wide variety of methods with the assumption that no one sampling method is suitable for all models of interest.
BCM \cite{thijssen2016bcm} offers both a generic interface (via C++) and a wide variety of samplers, but does not supply any likelihood functions and is unfortunately largely undocumented.

\begin{figure*}[h]
\centerline{\includegraphics[width=\textwidth]{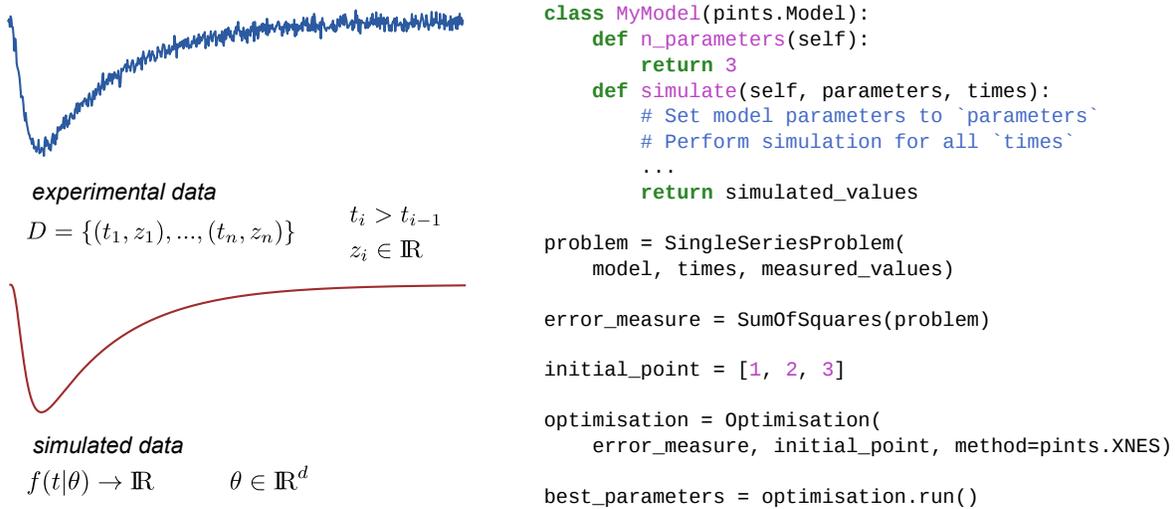}}
\caption{%
\emph{(Left)} An experimentally measured noisy time series, and a simulated one.
\emph{(Right)} An example of an optimisation procedure with PINTS.
Note that the actual simulation code is omitted from the example model wrapper at the top: this is the user-provided part, and can be written in Python or any other language that Python can interface with, allowing computationally heavy forward simulations to be handled entirely outside of PINTS.
}
\label{fig:overview}
\end{figure*}

\section*{Implementation and architecture}

PINTS is designed around two core ideas:
1. PINTS should work with a wide range of time series models, and make no demands on how they are implemented other than a minimal input/output interface.
2. It is assumed that model evaluation (simulation) is the most costly step in any optimisation or sampling routine.

The decision to use Python fits both these criteria: Python interfaces well with C and C++, which are typically used for high-performance simulation, and any performance hit of using the high-level, easy to read and write language Python is overshadowed by simulation time.


\paragraph{Defining an optimisation or sampling problem}

All optimisers operate on a callable \code{ErrorMeasure} object that describes a function to minimise, or on a callable \code{LogPDF} object that describes a probability density function (PDF) to maximise.
Similarly, all samplers start from a callable \code{LogPDF}, so that the same probability function can be used with both optimisers and samplers.
The natural logarithm of the PDF is used for computational efficiency and accuracy, and we assume that the probability density is unnormalised (i.e. its integral does not necessarily sum to 1).

\begin{figure*}[h]
\centerline{\includegraphics[width=\textwidth]{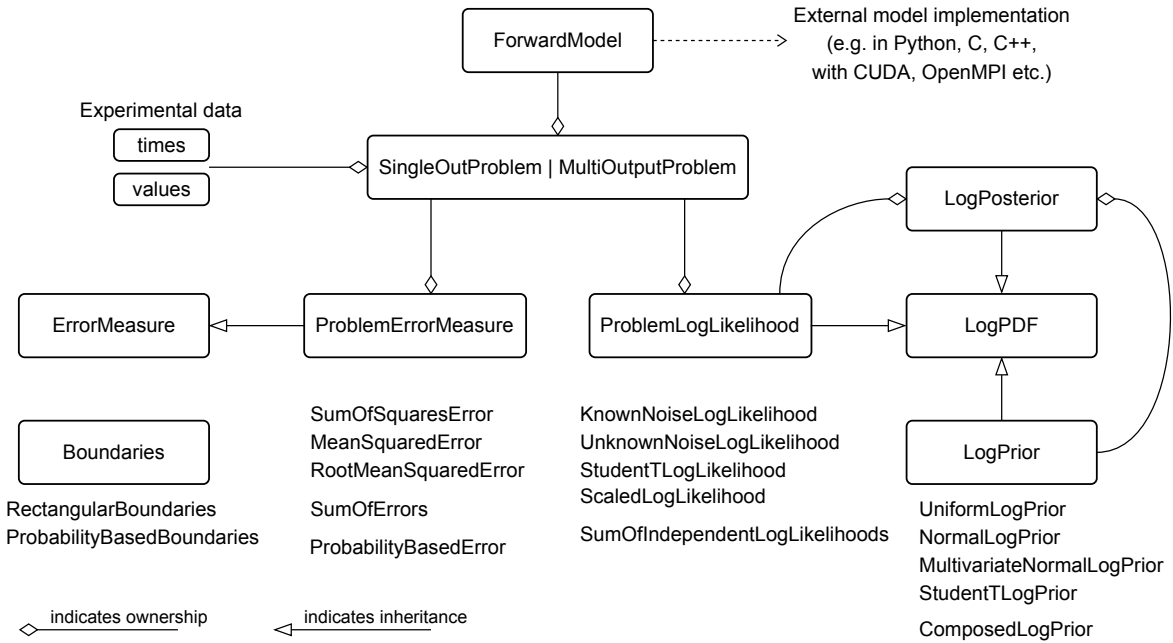}}
\caption{%
An overview of the PINTS classes used to define error measures (for optimisation) and PDFs (for optimisation or sampling).
Users can write a wrapper class for their model, making it available to pints, and must provide the experimental data using any Python sequence structure (for example, a list or a \code{NumPy} array).
With these ingredients, a (single or multi-output) problem can be defined that can then be used with any of the available error measure or likelihood classes.
Alternatively, users implement their own \code{ErrorMeasure} or \code{LogPDF}, which allows for further customisation and for problems other than time series problems to be solved.
}
\label{fig:uml}
\end{figure*}

Figure \ref{fig:uml} shows how a user-defined model can be wrapped in a PINTS \code{ForwardModel} and combined with time points and measured values to create a \code{Problem} from which several standard \code{ErrorMeasure}s and \code{LogPDF}s can be created.
For inference in a Bayesian context, a \code{LogPosterior} class and several \code{LogPrior} distributions are provided.
If a given \code{LogPDF} or \code{ErrorMeasure} cannot be constructed from PINTS classes, users can also define their own classes.

\paragraph{Implementation of optimisers and samplers}

\begin{figure*}[h]
\centerline{\includegraphics[width=0.8\textwidth]{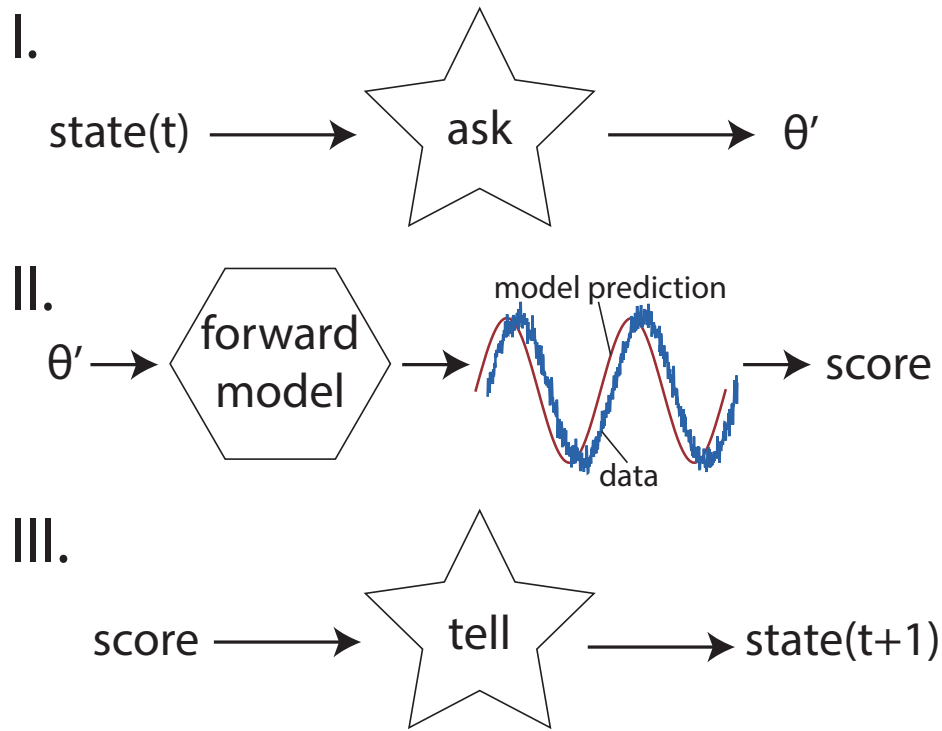}}
\caption{%
The three steps iterated in an ask-and-tell interface.
The stars here represent code specific to the chosen sampling or optimiser method and $\theta'$ is the input parameter vector proposed by \code{ask()}.
The \code{state(.)} of the system varies according to the method but typically holds a set of input parameter vectors and other constant or dynamic variables used by the \code{ask()} and \code{tell()} steps.
}
\label{fig:ask_tell}
\end{figure*}

Most PINTS samplers and optimisers are implemented using a so-called \emph{ask-and-tell interface}, inspired by the Python implementation of CMA-ES \cite{hansen2003reducing} (\url{https://github.com/CMA-ES/pycma}).
In this framework, the details of solving the forward problem are partitioned away from the rest of the sampling or optimising algorithm.
For each iteration the following steps are undertaken (Figure \ref{fig:ask_tell}):
first, the user calls \code{ask()} to obtain one or more parameter values from their chosen method --- these values are typically vectors generated stochastically conditional on an internal system state;
second, the user solves the forward model and generates a score for each parameter vector, for example, an error measure or (unnormalised) posterior probability;
third, the user calls \code{tell()} to pass the score back to the method, which can then update its internal state and finish the iteration.
For example, in many MCMC methods each \code{ask()} call returns a single proposed sample to be evaluated by the user, and the following \code{tell()} then either accepts or rejects this point based on its probability.
For optimisers such as CMA-ES, \code{ask()} returns a set of points in parameter space, and the scores passed in via \code{tell()} are then used to estimate the local gradient, which is used to move towards the estimated optimum.
This framework has a number of advantages:
since optimisation or sampling can take hours or even days, this allows programs using PINTS to provide regular user feedback and logging (which is not possible when the routine is implemented as a single monolithic function call); allows users with access to CPU clusters or GPU machines to implement their own parallelised evaluation of \code{ErrorMeasure}s and \code{LogPDF}s; lets users implement their own strategies (for example, by dynamically changing hyperparameters) and/or stopping criteria; and, finally, by delineating the sampling or optimisation algorithm's steps from the methods used to solve the forward problem, encourages development of transparent and modular code.

For more casual users (whom we expect will be the majority), PINTS provides standard \code{Optimisation} and \code{MCMCSampling} methods that can be run for a fixed (user-specified) number of iterations or, alternatively, until user-specified stopping criteria are reached.
These functions also allow logging and parallelised evaluations using Python multiprocessing.

\section*{Quality control}

PINTS has three levels of testing: unit testing, performance testing, and comparative testing.
\href{https://github.com/pints-team/pints/tree/master/pints/tests}{Unit tests} are used to test the functionality of simple (deterministic) methods, and to check that complex (pseudo-random) methods run without raising exceptions.
All the unit tests are available to be run by a user to ensure the software is working correctly.
Continuous integration is carried out using \href{https://travis-ci.org/pints-team/pints}{Travis CI} (Ubuntu Trusty distribution with Python versions 2.7, 3.4, 3.5 and 3.6, and OS/X with Python version 2.7) and \href{https://ci.appveyor.com/project/MichaelClerx/pints}{AppVeyor} (Windows Server 2011 R2 with Python versions 2.7, 3.4, 3.5 and 3.6).
PINTS uses Flake8 linter tests to ensure that contributed code conforms to best practices, and code coverage tests to ensure that all code is sufficiently tested and documented.
\href{https://github.com/pints-team/functional-testing}{Functional testing} of the methods is performed separately, and is used to test the method's behaviour from different (pseudo-random) initial conditions.
Analysis of functional checking is done both visually and statistically (for example, if recent results deviate significantly from previous results this indicates the possible introduction of a bug).
Finally, in \href{https://github.com/pints-team/performance-testing}{comparative testing} a number of problems are set up and solved with different methods, in order to compare the solutions they return and evaluate their performance.

\section*{(2) Availability}
\vspace{0.5cm}
\section*{Operating system}

PINTS uses no functions specific to any operating system (OS), and so can run on any OS that provides Python.
Optional parallelisation is provided that uses the Python \code{multiprocessing} module, which works best on UNIX-based systems (for example, Linux and OS/X), but runs on Windows with slightly reduced performance.

\section*{Programming language}

PINTS requires Python 2.7 or higher, or Python 3.4 or higher.

\section*{Additional system requirements}

PINTS has a minimal disk space footprint (approximately 2MB) and can be run on single-processor devices or headless on multi-processor machines (for example, via \code{ssh}).

\section*{Dependencies}

PINTS uses the NumPy (version 1.8 or higher) and SciPy (version 0.14 or higher, \cite{jones2014scipy}) libraries extensively.
The default optimisation method is CMA-ES, for which the \code{cma} package (version 2 or higher) is used.
The remaining optimisation and sampling methods require no further dependencies.
Finally, utility functions for plotting with Matplotlib (version 1.5 or higher \cite{hunter2007matplotlib}) are provided, but it is possible to use PINTS without Matplotlib or with a different plotting library.

\section*{List of contributors}


Michael Clerx,
Sanmitra Ghosh,
Ben Lambert,
Chon Lok Lei,
Martin Robinson.

\section*{Software location:}



\begin{description}[noitemsep,topsep=0pt]
  \item[Name:] GitHub (release v0.1.1)
  \item[Persistent identifier:] https://github.com/pints-team/pints/releases/tag/v0.1.1
  \item[Licence:] BSD 3-clause
  \item[Publisher:] Pints team
  \item[Version published:] 0.1.1
  \item[Date published:] 01/11/18
\end{description}

{\bf Code repository}

\begin{description}[noitemsep,topsep=0pt]
  \item[Name:] GitHub (develop)
  \item[Persistent identifier:] https://github.com/pints-team/pints
  \item[Licence:] BSD 3-clause
  \item[Date published:] \textcolor{blue}{16/05/17}
\end{description}

\section*{Language}

English.

\section*{(3) Reuse potential}


Detailed documentation is provided on using PINTS with user-supplied models (see, for example, the \code{writing-a-model} example on the GitHub repository).
While PINTS was designed primarily with biological and electrochemical problems in mind, there is nothing to prohibit its use on time series models from other fields.
Similarly, while the implemented \code{ErrorMeasure} and \code{LogPDF} classes were chosen to work well with time series problems, PINTS can be used outside this setting (in fact, utility functions \code{fmin} and \code{curve\_fit} are provided specifically for this purpose).
Users are also free to create their own \code{ErrorMeasure} or \code{LogPDF} classes which do not rely on PINTS \code{ForwardModel} or \code{Problem} classes.
A link to the full API documentation can be found on the GitHub repository, which also contains a list of examples for all PINTS' main features (\url{https://github.com/pints-team/pints/examples}).
We have found that these examples, rather than our API documentation, serve to kick-start any new project based on PINTS.
We welcome questions, bug reports, and user contributions via the same repository, which acts as a central communication platform for PINTS.

%

\section*{Funding statement}

M.C., G.R.M. and D.J.G. acknowledge support from the UK Biotechnology and Biological Sciences Research Council [BBSRC grant number BB/P010008/1];
M.R., S.G. and D.J.G. gratefully acknowledge research support from the UK Engineering and Physical Sciences Research Council Cross-Disciplinary Interface Programme [EPSRC grant number EP/I017909/1];
C.L. acknowledges support from the Clarendon Scholarship Fund, the EPSRC and the UK Medical Research Council (MRC) [EPSRC grant number EP/L016044/1];
BL acknowledges support from the UK Engineering and Physical Sciences Research Council [EPSRC grant number EP/F500394/1];
and S.G. and G.R.M. acknowledge support from the Wellcome Trust \& Royal Society [Wellcome Trust grant number 101222/Z/13/Z].

\section*{Competing interests}

The authors declare that they have no competing interests.




\bibliographystyle{agsm}
\bibliography{bibliography}

\vspace{2cm}

\rule{\textwidth}{1pt}

{ \bf Copyright Notice} \\
Authors who publish with this journal agree to the following terms: \\

Authors retain copyright and grant the journal right of first publication with the work simultaneously licensed under a  \href{http://creativecommons.org/licenses/by/3.0/}{Creative Commons Attribution License} that allows others to share the work with an acknowledgement of the work's authorship and initial publication in this journal. \\

Authors are able to enter into separate, additional contractual arrangements for the non-exclusive distribution of the journal's published version of the work (e.g., post it to an institutional repository or publish it in a book), with an acknowledgement of its initial publication in this journal. \\

By submitting this paper you agree to the terms of this Copyright Notice, which will apply to this submission if and when it is published by this journal.

\end{document}